\documentclass{kluwer}    
\usepackage{graphicx,psfig}

\newdisplay{guess}{Conjecture}

\begin{document}  
\begin{article}
\begin{opening}         
\title{X-ray, Ly\hspace{0.1mm}$\alpha$ and H$\alpha$ Emission from Simulated 
Disk Galaxies}
\author{Jesper \surname{Sommer--Larsen}}  
\institute{Theoretical Astrophysics Center, Copenhagen}
\author{Sune \surname{Toft}}  
\author{Jesper \surname{Rasmussen}}  
\author{Kristian \surname{Pedersen}}  
\institute{Astronomical Observatory, Copenhagen}
\author{Martin \surname{G\"otz}}  
\author{Laura \surname{Portinari}}  
\runningauthor{J.~Sommer-Larsen et~al.}
\institute{Theoretical Astrophysics Center, Copenhagen}
\runningtitle{X-ray, Ly\hspace{0.1mm}$\alpha$ and H$\alpha$ Emission from  
Simulated Disk Galaxies}

\begin{abstract}
The X-ray properties of the haloes of disk galaxies formed in fully
cosmological, hydro/gravity simulations are discussed. The results are
found to be consistent with observational X-ray detections and upper limits.
Disk galaxy haloes are predicted to be about an order of magnitude brighter
in soft X-rays at $z$~$\sim$~1 than at $z$=0.

The Ly\hspace{0.1mm}$\alpha$ and H$\alpha$ surface brightness of an edge-on,
Milky Way like model galaxy has been determined. The emission is found to
be quite extended, with a scale height of about 600 pc, neglecting extinction
corrections.

\end{abstract}
\end{opening}           
\vspace{-3mm}

\section{Introduction}  
\begin{figure}
\psfig{file=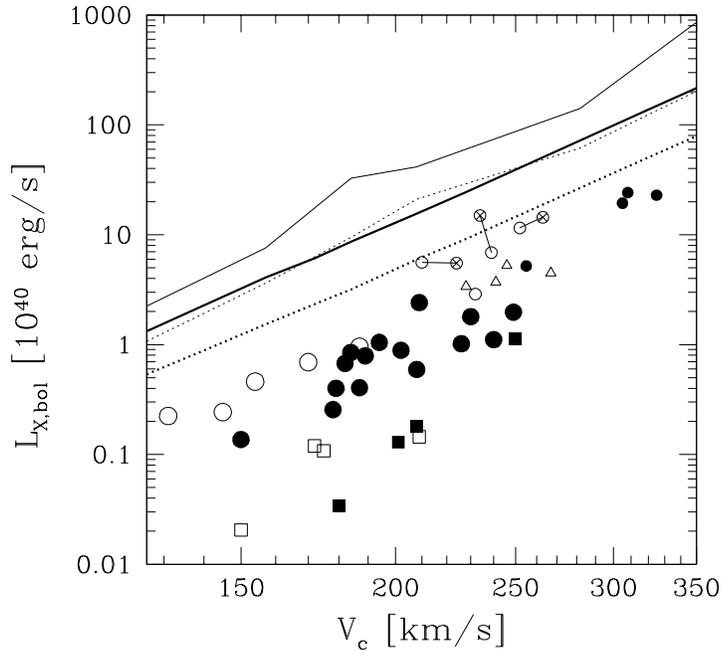,width=10truecm}
\caption{Bolometric luminosity at $z$=0 as a function of characteristic 
circular speed.
{\bf Small symbols}: Flat $\Omega_M=1.0$ cosmology:
Open symbols: baryon fraction $f_b$=0.05, filled circles
$f_b$=0.1. Triangles: without UV field, non-triangles: with a UV field
of the Efstathiou (1992) type. 
Connected symbols are the same galaxies run with medium (open
circles) and high (open circles with crosses) resolution. All  
simulations represented by small symbols have primordial abundance.
{\bf Large symbols}: Flat ($\Omega_{\Lambda},\Omega_M)=(0.7,0.3)$ cosmology: 
Open symbols: $f_b$=0.05, filled symbols $f_b$=0.1. Circles correspond to
primordial abundance and with a Haardt \& Madau (1996) UV field, squares
correspond to $Z=1/3~Z_{\odot}$ (using the cooling function of Sutherland \& 
Dopita 1993, which does not include effects of a UV field).
The curves are the $L_{X,bol}$-$V_c$ relationship for the simple cooling flow
models for $\Lambda$CDM NFW haloes --- see Toft et~al.\ (2002).
The curves represent different baryonic fractions
(solid curves have $f_b=0.1$, dotted curves have $f_b=0.05$) and
abundances (thick curves: primordial abundances, thin
curves: $Z=1/3~Z_{\odot}$).}
\end{figure}
\begin{figure}
\psfig{file=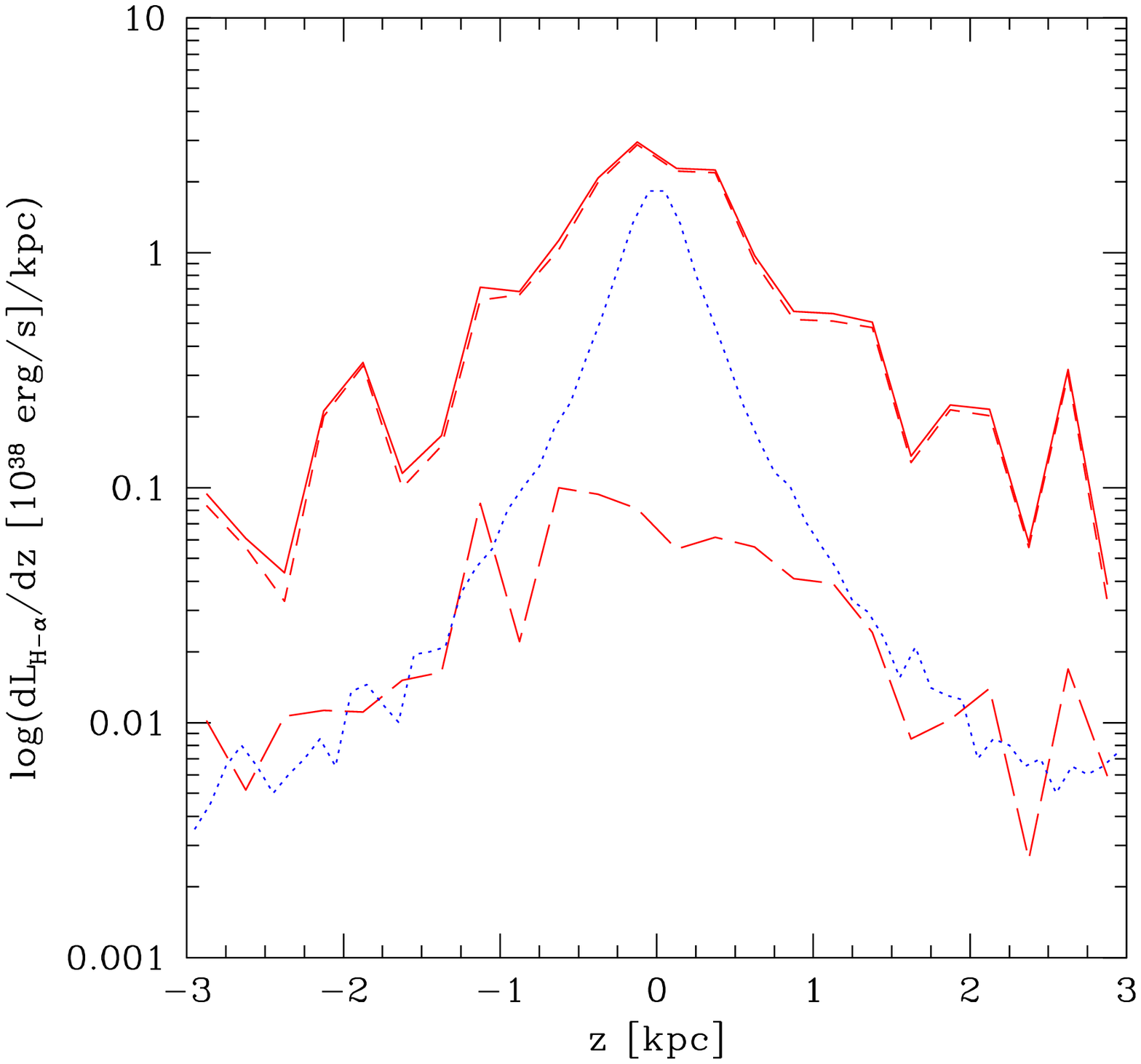,width=10truecm}
\caption{H$\alpha$ surface brightness of a Milky Way like, edge-on disk galaxy
at $z$=0 (solid curve ---
contribution from collisional excitation: short-dashed curve, recombination: 
long-dashed curve) - no extinction correction has been applied. Also shown is 
the surface density of cold ($\sim$10$^4$ K) gas (dotted curve).}
\end{figure}

\vspace{-3mm}
\noindent
Gradual
infall of halo gas onto the disk due to radiative cooling is a generic feature
of disk galaxy formation models. Such continuing gas infall seems
essential to explain the extended star formation histories of isolated spiral
galaxies like the Milky-Way and is the most likely explanation of the ``G-dwarf
problem'' --- see, e.g., Rocha-Pinto \& Maciel (1996) and Pagel (1997).

At the virial temperatures of disk galaxy haloes the dominant cooling
mechanism is thermal bremsstrahlung plus atomic line emission.
The emissivity, increasing strongly with halo gas density, is expected to 
peak fairly close to the disk and decrease outwards, 
and if the cooling rate is significant
the X-ray flux may be visible well beyond the optical radius of a galaxy.

Recently, Benson et~al.\ (2000) compared ROSAT observations 
of three
massive, nearby and highly inclined disk galaxies with predictions of simple
cooling flow models of galaxy formation and evolution.  
They showed that these models predict about {\emph{an order of
magnitude}} more X-ray emission from the galaxy haloes than observational 
detections and upper limits.

We have determined global X-ray properties of the haloes of a 
novel sample of 44 model disk galaxies at redshift $z$=0. The galaxies result 
from physically realistic, fully cosmological gravity/hydro simulations of 
galaxy formation and evolution. The galaxies span a range in
characteristic circular speeds of $V_c$ = 130--325 km s$^{-1}$ and have been
obtained
with a considerable range of physical parameters, varying the baryonic
fraction, the gas metallicity, the meta-galactic UV field, the cosmology, the
dark matter type, and also the numerical resolution. Details of the simulations
and halo X-ray emission calculations are given in Toft et~al. (2002), 
Sommer-Larsen \& Dolgov (2001) and Sommer-Larsen et~al.\ (2002).

\vspace{-3mm}

\section{X-ray emission}
\vspace{-4mm}
\noindent
In Fig. 1 the total bolometric X-ray luminosities $L_{X,bol}$ of the
44 simulated disk galaxies in our sample are plotted versus their 
characteristic circular speed $V_c$. Also shown are predictions by simple
cooling flow models. The X-ray luminosities derived from the 
simulations are up to two orders of magnitude
below values derived from simple models. Toft et~al.\ (2002) show that our 
model predictions of X-ray properties
of disk galaxy haloes are consistent with observational detections and
upper limits. As can be seen from the
figure $L_{X,bol}$ $\sim$
$10^{40}$ erg s$^{-1}$ for a Milky Way sized galaxy. This in turn implies that
hot halo gas is cooling out and being deposited onto the galactic disk at a
rate of {\mbox{$\sim$~0.5--1 $M_{\odot}$yr$^{-1}$}}, consistent with 
observational
upper limits, as discussed by  Sommer-Larsen et~al.\ (2002). They also show
that the present amount and distribution of hot gas in the haloes of Milky
Way like disk galaxies is consistent with observed dispersion measures towards
pulsars in the globular cluster M53 and the LMC. 

In contrast to what is predicted by simple cooling flow models, it is found
that {\it increasing} cooling efficiency of the halo gas results in
{\it decreasing} present day $L_X$.
The reason for this is that increasing the cooling efficiency over
the course of a simulation results in less hot gas in the halo at $z$=0
to cool (because the total amount of gas available at any given time
is always limited to the gas inside of the virial radius). This in turn leads 
to lower  present day accretion rates and lower $L_{X,bol}$.

Finally, it is found for realistic choices of the physical parameters
that disk galaxy haloes were up to {\it one order of magnitude}
brighter in soft X-ray emission at $z$$\sim$1, than at present.

\vspace{-3mm}
\section{Ly\hspace{0.1mm}$\alpha$ and H$\alpha$ emission}
\vspace{-4mm}
In the simulations, the local ionization balance of Hydrogen is assumed to be 
set by collisional ionization, photo-ionization by the redshift dependent, 
meta-galactic UV field
and recombination. Radiative transfer of the ionizing UV photons is included
in a simplified way. Given this, the local rate of Ly\hspace{0.1mm}$\alpha$ 
and H$\alpha$ emission can be calculated. Fig. 2 shows the appearance of an
edge-on, Milky Way like disk galaxy in H$\alpha$ (Ly\hspace{0.1mm}$\alpha$ is
very similar). No extinction correction has been applied; such a correction
will clearly be very important within 100-200 pc from the midplane of the 
disk. The distribution of  
H$\alpha$ emission is considerably more extended than that of the cold gas
and (since the H$\alpha$ emission scale height is about 600 pc) also of the
stars. This is in broad agreement with observations (e.g., Hoopes et~al.\
1999, Wang et~al.\ 2001, Olsen et~al.\ 2002).

\vspace{-3mm}
\section*{Acknowledgement}
\vspace{-3mm}
We thank the organizers for a splendid conference.

\noindent

\vspace{-1mm}


\end{article}

\end{document}